\documentclass[prl,twocolumn,showpacs,superscriptaddress]{revtex4}
\usepackage[dvips]{graphicx}
\usepackage{hyperref}
\usepackage{bm}
\usepackage{times}

\begin{document}

\def\jpb{J. Phys. B: At. Mol. Opt. Phys.~}
\def\pra{Phys. Rev. A~}
\def\prb{Phys. Rev. B~}
\def\prl{Phys. Rev. Lett.~}
\def\jmo{J. Mod. Opt.~}
\def\jetp{Sov. Phys. JETP~}
\def\etal{{\em et al.}}

\def\vekt#1{{\bm{#1}}}
\def\vect#1{\vekt{#1}}
\def\vektalpha{\vekt{\alpha}}
\def\vektr{\vekt{r}}
\def\vektp{\vekt{p}}
\def\vektpc{\vekt{p}_{\mathrm{c}}}
\def\vektE{\vekt{E}}
\def\vekte{\vekt{e}}
\def\vektA{\vekt{A}}
\def\vektEhat{\hat{\vekt{E}}}
\def\vektB{\vekt{B}}
\def\vektv{\vekt{v}}
\def\vektk{\vekt{k}}
\def\vektkhat{\hat{\vekt{k}}}
\def\reff#1{(\ref{#1})}

\def\Up{U_{\mathrm{p}}}
\def\Ip{I_{\mathrm{p}}}
\def\Tp{T_{\mathrm{p}}}
\def\Ss{S_{\vektp\mathrm{s}}}
\def\SIs{S_{\vektp\Ip\mathrm{s}}}
\def\Cps{C_{\vektp\mathrm{s}}}
\def\Cpnulls{C_{\vektp_0\mathrm{s}}}
\def\SI{S_{\vektp\Ip}}
\def\SIsnull{S_{\vektp_0\Ip\mathrm{s}}}
\def\zt{z_{\mathrm{t}}}
\def\ts{t_{\vektp\mathrm{s}}}
\def\tsnull{t_{\vektp_0\mathrm{s}}}
\def\tnulltilde{\tilde{t}_0}
\def\omegap{\omega_p}
\def\omegaMie{\omega_M}
\def\diff{\mathrm{d}}
\def\imagi{\mathrm{i}}
\def\eulere{\mathrm{e}}

\def\halb{\frac{1}{2}}

\def\beq{\begin{equation}}
\def\eeq{\end{equation}}

\def\energy{{\cal E}}

\def\Ehat{\hat{E}}
\def\Ahat{\hat{A}}

\def\ket#1{\vert #1\rangle}
\def\bra#1{\langle#1\vert}
\def\braket#1#2{\langle #1 \vert #2 \rangle}

\def\makered#1{{\color{red} #1}}

\def\Re{\,\mathrm{Re}\,}
\def\Im{\,\mathrm{Im}\,}

\def\varphic{\varphi_{\mathrm{c}}}

\def\vxc{v_\mathrm{xc}}
\def\vextop{\hat{v}_\mathrm{ext}}
\def\VC{V_\mathrm{c}}
\def\VHX{V_\mathrm{Hx}}
\def\VHXC{V_\mathrm{Hxc}}
\def\wop{\hat{w}}
\def\Gammaevenodd{\Gamma^\mathrm{even,odd}}
\def\Gammaeven{\Gamma^\mathrm{even}}
\def\Gammaodd{\Gamma^\mathrm{odd}}

\def\Hop{\hat{H}}
\def\HopKS{\hat{H}_\mathrm{KS}}
\def\HKS{H_\mathrm{KS}}
\def\Top{\hat{T}}
\def\TopKS{\hat{T}_\mathrm{KS}}
\def\VopKS{\hat{V}_\mathrm{KS}}
\def\VKS{{V}_\mathrm{KS}}
\def\vKS{{v}_\mathrm{KS}}
\def\Ttildeop{\hat{\tilde{T}}}
\def\Ttilde{{\tilde{T}}}
\def\Vextop{\hat{V}_{\mathrm{ext}}}
\def\Vext{V_{\mathrm{ext}}}
\def\Vopee{\hat{V}_{{ee}}}
\def\psiopdag{\hat{\psi}^{\dagger}}
\def\psiop{\hat{\psi}}
\def\vext{v_{\mathrm{ext}}}
\def\Vee{V_{ee}}
\def\vee{v_{ee}}
\def\nop{\hat{n}}
\def\Uop{\hat{U}}
\def\Wop{\hat{W}}
\def\bop{\hat{b}}
\def\bopdag{\hat{b}^{\dagger}}
\def\qop{\hat{q}}
\def\jop{\hat{j\,}}
\def\vHxc{v_{\mathrm{Hxc}}}
\def\vHx{v_{\mathrm{Hx}}}
\def\vH{v_{\mathrm{H}}}
\def\vc{v_{\mathrm{c}}}
\def\xop{\hat{x}}

\def\Wcmcm{W/cm$^2$}

\def\varphiexact{\varphi_{\mathrm{exact}}}

\def\fmathbox#1{\fbox{$\displaystyle #1$}}

\title{Low-Energy Structures in Strong Field Ionization Revealed by Quantum Orbits}

\author{Tian-Min Yan}
\affiliation{Institut f\"ur Physik, Universit\"at Rostock, 18051 Rostock, Germany}
\affiliation{Max-Planck-Institut f\"ur Kernphysik, Postfach 103980, 69029 Heidelberg, Germany}

\author{S.V.~Popruzhenko}
\affiliation{National Research Nuclear University ``Moscow Engineering Physics Institute'', Kashirskoe Shosse 31, 115409, Moscow, Russia}

\author{M.J.J.~Vrakking}
\affiliation{FOM-Institute AMOLF, Science Park 113, 1098 XG Amsterdam, The Netherlands}
\affiliation{Max-Born-Institut, Max-Born-Stra{\ss}e 2A, 12489 Berlin, Germany}

\author{D.~Bauer}
\thanks{Corresponding author: dieter.bauer@uni-rostock.de}
\affiliation{Institut f\"ur Physik, Universit\"at Rostock, 18051 Rostock, Germany}

\date{\today}

\begin{abstract} Experiments on atoms in intense laser pulses and the corresponding exact {\em ab initio} solutions of the time-dependent Schr\"odinger equation (TDSE) yield photoelectron spectra with low-energy features that are not reproduced  by the otherwise successful work horse of strong field laser physics: the ``strong field approximation'' (SFA). In the semi-classical limit, the SFA possesses an appealing interpretation in terms of interfering quantum trajectories. It is shown that a conceptually simple extension towards the inclusion of Coulomb effects yields very good agreement with exact TDSE results. Moreover, the Coulomb quantum orbits allow for a physically intuitive interpretation and detailed analysis of all low-energy features in the semi-classical regime, in particular the recently discovered ``low-energy structure'' [C.I.~Blaga {\em et al.}, Nature Physics {\bf 5}, 335 (2009) and W.\ Quan {\em et al.}, Phys.\ Rev.\ Lett.\ {\bf 103}, 093001 (2009)].          
\end{abstract}
\maketitle

The development of analytical and numerical methods capable of treating strongly-driven quantum systems is of great interest in many areas of physics. By ``strongly-driven'' we understand that conventional time-dependent perturbation theory is not applicable. A prime example for such a system is an atom in an intense laser field. The force on valence electrons due to the electric field of the electromagnetic wave delivered by present-day intense lasers can easily compete with the binding force.  As a consequence,  the photoelectron spectra may show strong nonperturbative features such as plateaus and cut-offs \cite{milo06}, instead of a simple exponential decrease with the number of absorbed photons, as expected from perturbation theory. Recently, an ``ionization surprise'' \cite{fais09} at wavelength $\lambda=2\,\mu$m and intensity $I=80$--$150$\,TW/cm$^2$, the so-called ``low-energy structure'' (LES) \cite{blaga09,quan09}, has been reported. The LES is a strong but narrow enhancement of the differential ionization probability along the polarization direction of the laser at low energies. This result was so astonishing not only because it is unpredicted by the  ``strong field approximation'' (SFA) \cite{kfr} but also because it is observed in a regime where matters were actually expected to simplify. In fact, if the number of photons $N$ of energy $\hbar\omega$ required to overcome the ionization potential $\Ip$ is large, $N=\Ip/\hbar\omega\gg 1$, and the time the electron needs to tunnel through the Coulomb-barrier is small compared to a laser period, i.e., the Keldysh parameter $\gamma=\sqrt{\Ip/2\Up}$ with $\Up$ the ponderomotive potential, is small, a quasi-static tunneling theory appears to apply \cite{popov04}. As the tunneling ionization rate in a static electric field is a smooth, featureless function of the final momenta $p_\parallel$ and $p_\perp$ parallel and perpendicular to the electric field, respectively, no LES has been expected. In the present Letter we reveal the origin of the LES using our trajectory-based Coulomb-SFA (TC-SFA). The fact that the TC-SFA allows recourse to trajectories provides an unprecedented insight into the origin of any spectral feature of interest, as constructive or destructive interference of trajectories or the Coulomb-focusing of them \cite{brab} can be analyzed in all details. 

\begin{figure}
\includegraphics[width=0.4\textwidth]{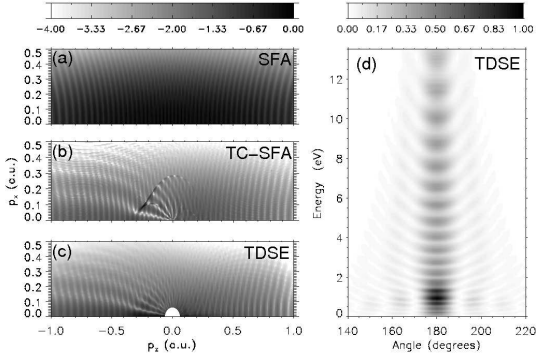} 
\caption{Logarithmically scaled photoelectron momentum distributions in the $p_zp_x$-plane ($p_\parallel=p_z$) calculated using (a) plain SFA, (b) TC-SFA, and (c) numerical solution of the TDSE. Panel (d) shows the linearly scaled, angle-resolved energy spectrum in backward-direction with the LES around $1$\,eV clearly visible.  The laser and atomic parameters are given in the text. Each spectrum was normalized to its maximum value. The small empty semi-circle around the origin in the TDSE result (c) is due to the mapping of the angle-resolved energy spectra (as calculated by Qprop \cite{qprop}) to the Cartesian $p_zp_x$-momentum plane.
     \label{fig1}}
\end{figure}

The SFA  is a widely and successfully used theoretical approach to tackle strong field photo detachment. In its simplest form it accounts only for the so-called ``direct'' electrons, which are bound to the atom up to the detachment time $t_0$ and at later times just move in the laser field without any interaction with the parent atom anymore. Quantum mechanically, such a free electron with drift-momentum $\vektp$ in a laser field defined by the vector potential $\vektA(t)$ is described in length gauge by a Gordon-Volkov state \cite{gv} $\ket{\Psi^{\mathrm{(GV)}}_\vektp(t)}=\eulere^{-\imagi S_\vektp(t)}\ket{\vektp+\vektA(t)}$, $S_\vektp(t)=\int^t [\vektp+\vektA(t')]^2/2\,\diff t'$ (atomic units are used unless noted otherwise). The SFA-transition matrix element for a final asymptotic momentum $\vektp$ (at the detector) for a laser pulse $E(t)=-\partial_t A(t)$ in dipole approximation, polarized along the $z$-axis, and lasting from  $t=0$ until $t=\Tp$ then reads 
\beq M^{\mathrm{(SFA)}}_\vektp=-\imagi \int_0^{\Tp} \bra{\Psi^{\mathrm{(GV)}}_\vektp(t)} zE(t) \ket{\Psi_0(t)} \,\diff t \label{SFA1} \eeq
where $\ket{\Psi_0(t)}=\eulere^{\imagi \Ip t}\ket{\Psi_0}$, and $\ket{\Psi_0}$ is the initial bound state of the electron. Alternatively, we may use the matrix element $\tilde{M}^{\mathrm{(SFA)}}_\vektp = \eulere^{-\imagi S_\vektp(\Tp)-\imagi \Ip \Tp}  M^{\mathrm{(SFA)}}_\vektp$, as it gives the same photoelectron spectrum, $\diff w(\vektp)/\diff^3 p=\vert M^{\mathrm{(SFA)}}_\vektp\vert^2 = \vert \tilde{M}^{\mathrm{(SFA)}}_\vektp\vert^2$. The neglect of any further interaction between electron and binding potential once the electron is ejected is well-justified for short-range potentials as, e.g., in the photodetachment from negative ions \cite{manakov09,gazi}.  However, the agreement with photoelectron spectra calculated {\em ab initio} by solving the TDSE is in general poor for the case of long-range binding potentials, such as in the more common ionization of neutral atoms or positive ions \cite{milo06,arbo}. For illustration, we show in Fig.~\ref{fig1} SFA and TDSE photoelectron momentum spectra for a linearly polarized $n=3$-cycle pulse of the form
\beq \vektA(t)=-\frac{\Ehat}{\omega}\vekte_z \sin^2\left(\frac{\omega t}{2n}\right)  \sin\omega t \label{vecpotpulse} \eeq
for $t\in[0,2n\pi/\omega]$, and zero otherwise. The peak field strength $\Ehat=0.0534$ corresponds to $100$\,TW/cm$^2$, and the laser frequency $\omega=0.0228$ to $\lambda=2\,\mu$m. The binding potential used in the TDSE solver Qprop \cite{qprop} was  $V(r)=-1/r-17.0 \eulere^{-17.43 r}/r$, which yields $\Ip=0.579$ for the 1s state and a grid spacing $\Delta r= 0.2$. The same $\Ip$ and a Coulombic 1s initial state was used in the SFA-calculation. The particular choice of $\Ip$ was motivated by the experiment in \cite{blaga09,cat09} where argon was used.  Figure \ref{fig1} shows several striking discrepancies between SFA and TDSE: (i) the SFA predicts a symmetric momentum distribution for a pulse of the form \reff{vecpotpulse} with an integer number of cycles $n$ while the TDSE spectrum is strongly asymmetric \cite{chelband}. (ii) The radial structures \cite{arbo,gopal} around $\vektp=\vekt{0}$ present in the TDSE spectrum are absent in the SFA result. (iii) Several side-lobes in the TDSE result are clearly visible for $p_z<0$ but completely absent in the SFA. 
In Fig.~\ref{fig1}b we anticipate the spectrum obtained with our TC-SFA method, which is---as regards points (i)--(iii) above---in excellent agreement with the TDSE result. Only the probability along a clearly visible caustic is overestimated. However, the classical caustic turns out to be related to the LES, as will be discussed below.

The TDSE momentum spectrum in Fig.~\ref{fig1}c does not allow a clear identification of the LES. In order to compare the TDSE results with Refs.~\cite{blaga09,cat09} angle-resolved energy spectra $\diff w(\energy,\theta)/ \sin\theta \diff \theta \diff \energy =p\vert M_\vektp\vert^2$ (where $\energy=p^2/2$ and $\tan\theta = \vert\vektp_\perp\vert/p_\parallel$) were calculated. Figure~\ref{fig1}d clearly shows the LES for the directional energy spectrum anti-parallel ($\theta=180^\circ$) to the polarization axis. Because of the short pulse duration, no LES is visible for $\theta=0$ (not shown), which helps to identify unambiguously the origin of the LES in the following. Figure~\ref{fig1}d compares well with the experimental result in Fig.~3b of \cite{cat09}.

Let us now introduce the TC-SFA approach. The time integral in \reff{SFA1} can be evaluated semi-classically for sufficiently big $N$ with the help of the saddle-point approximation leading to 
\beq \tilde{M}^{\mathrm{(SFA)}}_\vektp \simeq \sum_s  \Cps \eulere^{-\imagi \SIs} \eeq
where $ \SIs=\SI(\ts)$, $ \SI(t)=\int_{t}^{\Tp} \{  [\vektp+\vektA(t)]^2/2 $ $+ \Ip \}\diff t $ and $\Cps$ is a prefactor \cite{milo06,pop08}.
Here, the sum is over all saddle-point times $\ts$ fulfilling the stationary action equation 
\beq \left.\frac{\partial \SI}{\partial t}\right|_{\ts}=0 \ \Leftrightarrow \ \halb [\vektp+\vektA(\ts)]^2  = -\Ip.  \label{saeq}\eeq
For a given final momentum $\vektp$ there will, in general, be two complex saddle-point times $\ts$ per laser cycle. The saddle-point action $\SIs$ can be separated into  an integral from the complex saddle-point time $\ts$ down to the real axis where $t=\Re \ts\equiv t_0$ and then along the real axis from $t_0$ to $\Tp$, i.e., $\SIs=\SIs^{\downarrow} + \SIs^{\to}$. In plain SFA both contributions can be calculated analytically for a given (analytic) vector potential. However, the TC-SFA requires the numerical evaluation of electron trajectories in the presence of both the binding potential and the laser field as well as the corresponding  modified action along the real time axis $\SIs^{\mathrm{C}\to}$. Hence it is illustrative to first inspect such trajectories for the plain SFA case. Consider the equations of motion $\dot{\vektr}=\vektv=\vektp+\vektA(t)$ and $\dot{\vektp}=\vekt{0}$. Clearly, $\vektp=$const., and for a given saddle-point time $\ts$ the real part of the action reads $\SIs^{\to}=\int_{t_0}^{\Tp} \left\{\vektv^2/2 + \Ip\right\}\,\diff t $, as required. If we actually want to solve for $\vektr(t)$ we need to fix the initial conditions for $\vektv$ and $\vektr$. The initial condition for momentum (or velocity) follows from \reff{saeq}. For the initial position we choose  $\Re \vektr(\ts)=\vekt{0}$, which implies $\vektr(t_0)=\vektalpha(t_0)-\Re\vektalpha(\ts)$ with $\vektalpha(t)=\int^t\vektA(t')\,\diff t'$. In the tunneling-limit $\vektr(t_0)$ can be identified with the geometrical tunnel exit $\zt(t_0)$ given by $-\Ip = E(t_0)\zt(t_0)$.   

In our TC-SFA the equation of motion $\dot{\vektp}=\vekt{0}$ is replaced by $\dot{\vektp}=-\vektr/|\vektr|^3$, and simultaneously $\vektp$ is replaced by $\vektp_0$ in  \reff{saeq}, leading to a modified saddle-point time $\tsnull$ (and real part $\tilde{t}_0\equiv\Re \tsnull$). The TC-SFA saddle-point action reads $\SIsnull^{\mathrm{C}}=\SIsnull^{\mathrm{C}\downarrow} + \SIsnull^{\mathrm{C}\to}$ with $\SIsnull^{\mathrm{C}\downarrow}=\int_{\tsnull}^{\tnulltilde} \left\{ [\vektp_0+\vektA(t)]^2/2 + \Ip\right\}\diff t$ and  $\SIsnull^{\mathrm{C}\to}=\int_{\tnulltilde}^{\Tp} \left\{ \vektv^2(t)/2 - 1/r(t) + \Ip \right\}\diff t$. The action of the Coulomb-potential on the imaginary dynamics during the tunneling process affects the ionization probability \cite{popov04,pop08} but does hardly change the shape of the photoelectron spectra so that we neglect it in this work. However, note that due to the replacement of $\vektp$ by $\vektp_0$ and the Coulomb-modified saddle-point times nevertheless $\SIsnull^{\mathrm{C}\downarrow}\neq \SIs^{\downarrow}$. 

The TC-SFA ``shooting''-method is implemented as follows. Two loops run over $p_{0z}\in [-p^{\max}_{0z},p^{\max}_{0z}]$ and $p_{0x}\in [-p^{\max}_{0x},p^{\max}_{0x}]$. Because of cylindrical symmetry about $p_\parallel=p_z$ sampling of the $p_{0z}p_{0x}$-plane is sufficient. For each initial momentum Eq.~\reff{saeq} is solved for $\tsnull$ using a complex-root-finding routine. Each $\tsnull$ corresponds to a trajectory. Hence, there is another loop over all $\tsnull$ found. The complex part  $\SIsnull^{\mathrm{C}\downarrow} $ of the action for a given $\tsnull$ is calculated analytically, neglecting Coulomb effects. The trajectory corresponding to a $\tsnull$ is calculated from $\tnulltilde$ up to the end of the pulse according to the equations of motion $\dot{z}=p_z+A$, $\dot{x}=p_x$, $\dot{p}_z=-z/(x^2+z^2)^{3/2}$, $\dot{p}_x=-x/(x^2+z^2)^{3/2}$ (all time-arguments suppressed) for the initial conditions $x(\tnulltilde)=0$, $z(\tnulltilde)=\alpha_z(\tnulltilde)-\Re\alpha_z(\tsnull)$, $p_z(\tnulltilde)=p_{0z}$, $p_x(\tnulltilde)=p_{0x}$ using a Runge-Kutta solver. If the energy $\energy(\Tp)=p^2(\Tp)/2-1/r(\Tp)$ is negative, the trajectory does not correspond to a free electron and thus does not contribute to the photoelectron spectrum \cite{nubb,shv}. If the energy is positive the asymptotic momentum $\vektp$ can be calculated from $\vektp(\Tp)$ and $\vektr(\Tp)$  using Kepler's laws \cite{arbo07}, avoiding unnecessary explicit propagation up to large times.    Finally, the result for the $\tsnull$ under consideration is stored in a table of the form $p_{0z}$, $p_{0x}$, $p_z$, $p_x$, $\tsnull$, $\SIsnull^{\mathrm{C}}$, $\ldots$. Once the loops over the initial momenta are completed, the table with the trajectory data can be post-processed. The trajectories are binned according to their asymptotic momentum, and the TC-SFA matrix element
\beq \tilde{M}^{\mathrm{(TC)}}_\vektp = \sum_s \Cpnulls {\eulere^{-\imagi \SIsnull^{\mathrm{C}} }} \label{tcsfa}\eeq
is calculated, where the sum is over all trajectories ending up in the final momentum bin centered at $\vektp$.

The TC-SFA result in Fig.~\ref{fig1}b was obtained with $\simeq 2\times 10^8$ trajectories and shows the above introduced Coulomb-features (i)--(iii) of the TDSE calculation. In addition there is a caustic structure with a maximum at $\vektpc=(p_{\mathrm{c}\parallel},p_{\mathrm{c}\perp})=(-0.22,0.1)$. We have checked that this maximum moves closer towards the polarization axis as the wavelength is increased (with all other laser and target parameters held constant). For shorter wavelengths it fades away while becoming more ring-like. This is a strong indication that the LES of the full quantum TDSE simulations manifests itself as a caustic in the semi-classical TC-SFA approach.

It was argued already in the accompanying article \cite{fais09} that the LES is due to low-energy forward scattering  at the Coulomb-potential. Certainly, only electrons emitted with high probability, i.e., when the absolute value of the electric field is high, can contribute to  such a pronounced spectral feature as the LES. The quantum trajectories of the TC-SFA responsible for the caustic illustrate and confirm this view-point. For further analysis, we group all trajectories into four types. The first two types are the so-called ``short'' and ``long trajectories'' of plain SFA.  The short trajectory (type I) fulfills $\zt p_z > 0$ and $p_x p_{0x}>0$, meaning that the ejected electron moves from the tunnel exit directly towards the detector and has no close encounters with the ion. Coulomb effects on this type of trajectory are expected to be small. Type II is a ``long trajectory'' obeying  $\zt p_z < 0$ and $p_x p_{0x}>0$. Here, the electron starts from the tunnel exit that points away from the detector but ends up with a parallel momentum in the opposite direction because of the drift it acquires from the laser field at the time of emission, i.e., in plain SFA $p_z=p_{0z}=-A(t_0)$. The lateral initial momentum $p_{0x}$ of type-I and type-II trajectories is already in the same direction as the final momentum at the detector. This is always the case in plain SFA, as $p_x=p_{0x}$. Hence there are only type-I and type-II trajectories in plain SFA. While in Ref.~\cite{pop08} the type-I and type-II trajectories existing already in plain SFA are Coulomb-corrected, the shooting-method introduced in the present  Letter goes beyond a correction by taking into account qualitatively new types of trajectories generated by the binding potential. Type III is classified via $\zt p_z < 0$ and $p_x p_{0x}<0$. The corresponding electrons start on the opposite tunnel exit, as type-II electrons do, but moreover have an initial lateral momentum pointing in the ``wrong'' direction. Only due to relatively close encounters with the ion, the lateral momentum is ultimately reversed.  Finally, type-IV trajectories obey $\zt p_z > 0$ and $p_x p_{0x}<0$.

\begin{figure}
\includegraphics[width=0.45\textwidth]{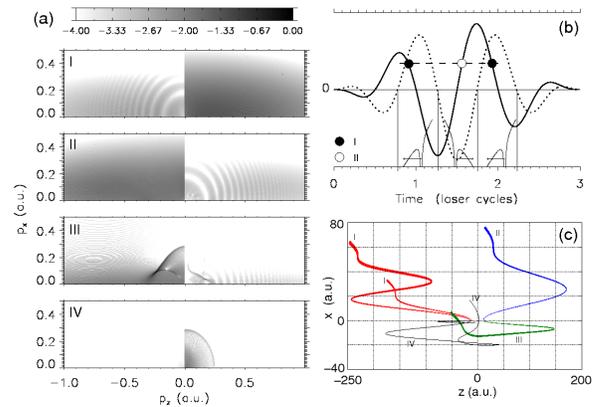} 
\caption{(color online). (a) Partial spectra due to trajectories I--IV. Panel (b) shows $A(t)$ (solid) and $E(t)$ (dotted) and illustrates why there is an interference structure for $p_z<0$ in (a) for type-I trajectories but not for type II. (c) The dominant trajectories in the $zx$-plane up to the time $\Tp$ which contribute to the final momentum $\vektpc$, i.e.,  where the caustic structure is visible for the type-III spectrum in panel (a). \label{fig3}}
\end{figure}

In order to understand the origin of spectral features such as the LES it is useful to study the partial contributions of each class of trajectories to the total spectrum in Fig.~\ref{fig1}b. To that end we split the sum over all trajectories in \reff{tcsfa} into four terms, $\tilde{M}^{\mathrm{(TC)}}_\vektp=\sum_{\nu=\mathrm{I}}^{\mathrm{IV}}\tilde{M}^{\mathrm{(TC)}}_{\nu\vektp}$, where the lower index $\nu$ indicates the trajectory type. Figure~\ref{fig3}a shows the partial spectra $\vert\tilde{M}^{\mathrm{(TC)}}_{\nu\vektp}\vert^2$. The spectrum  $\vert\tilde{M}^{\mathrm{(TC)}}_{\mathrm{I}\vektp}\vert^2$ allows only trajectories of type I to interfere with each other. Figure~\ref{fig3}b illustrates why there is only an interference pattern for $p_z<0$ but none for $p_z>0$: because of $p_z\simeq -A(t_0)$ the vector potential $A$ at the time of emission for $p_z<0$ is positive. For a positive $A$ there are two contributions (filled black circles in Fig.~\ref{fig3}b). The tunnel exit at those two times is at $z<0$, so that indeed the two trajectories are both of type I.  The open circle in   Fig.~\ref{fig3}b indicates a type-II trajectory because $\zt>0$. There is just one dominant type-II trajectory, which explains that there is no interference pattern in the spectrum $\vert\tilde{M}^{\mathrm{(TC)}}_{\mathrm{II}\vektp}\vert^2$ for $p_z<0$ in  Fig.~\ref{fig3}a. Figure~\ref{fig3}c shows  the dominant trajectories in the $zx$-plane which contribute to the final momentum $\vektpc$, i.e.,  where the caustic structure is visible for the type-III spectrum in panel  Fig.~\ref{fig3}a. As is evident from Fig.~\ref{fig3}c, the type-III electrons are driven back close to the origin and slowed down by the laser field.

\begin{figure}
\includegraphics[width=0.4\textwidth]{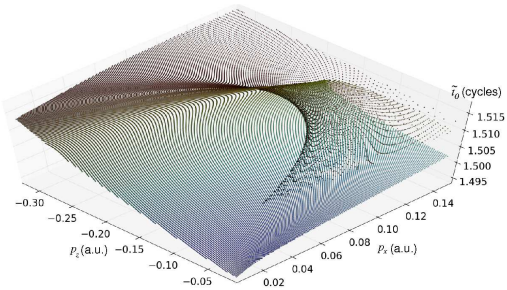} 
\caption{(color online). Emission times of type-III trajectories contributing to final momenta where the caustic structure is located. At $\vektpc$ the number of contributing trajectories is particularly high (steep slope).  \label{fig4}}
\end{figure}

We finally analyze how the caustic structure in the spectrum  $\vert\tilde{M}^{\mathrm{(TC)}}_{\mathrm{III}\vektp}\vert^2$ of Fig.~\ref{fig3}a is formed. To that end we added all type-III contributions incoherently and obtained a spectrum very similar to $\vert\tilde{M}^{\mathrm{(TC)}}_{\mathrm{III}\vektp}\vert^2$ of Fig.~\ref{fig3}a. Hence, interference can be ruled out, and, indeed, the LES has been observed in classical ensemble calculations as well \cite{quan09}. Our TC-SFA simulation reveals that the caustic structure is due to an enhancement of the number of trajectories of type III. This is shown explicitly in Fig.~\ref{fig4} where the emission times $\tilde{t}_0$ of type-III trajectories contributing to certain final momenta around the caustic structure are indicated. Particularly many type-III trajectories contribute at $\vektpc$, forming the semi-classical analogue of the LES: a caustic. Analytically, the mapping $\{p_{0x},p_{0z}\}\to\{p_x,p_z\}$ has a singularity at the caustic. In the numerical TC-SFA calculations this singularity manifests itself as an increase of the number of trajectories ending up close to $\vektpc$. The same phenomenon is known for rescattering where the trajectories merge and branch at classical cut-offs \cite{gor99}. Corresponding exact quantum mechanical calculations show an enhancement of the probability there \cite{carla04,manakov09}. As is visible in Fig.~\ref{fig3}a, panel III, the LES-caustic is a classical cut-off for type-III trajectories with not too small asymptotic lateral momenta. Feature (ii) from above, i.e., the radial structures around $\vektp=\vekt{0}$, are due to interference of all four types of trajectories. The side-lobes for $p_z<0$ (iii) emerge due to interference of type-II and type-III trajectories \cite{huis}. Both features  are not reproducible by purely classical ensemble simulations but require quantum approaches including Coulomb effects.

In summary, we presented a semi-classical approach based on quantum orbits to calculate the strong field ionization matrix element including Coulomb effects. The method yields an unprecedented agreement with the full {\em ab initio} solutions of the time-dependent Schr\"odinger equation. The method allows to analyze any spectral feature in terms of interfering or Coulomb-focused quantum trajectories. In particular, the recently observed low-energy structure at long wavelengths was found to originate from low-energy forward scattering, leading to caustic structures in semi-classical trajectory-calculations.

The authors acknowledge fruitful discussions with W.\ Becker, S.P.\ Goreslavski, J.-M.\ Rost, and N.I.\ Shvetsov-Shilovski. The work was partially supported by the Deutsche Forschungsgemeinschaft and the Russian Foundation for Basic Research. TMY acknowledges support from the International Max Planck Research School for Quantum Dynamics in  Heidelberg. The work of MJJV is part of the research program of the ``Stichting voor Fundamenteel Onderzoek der
Materie (FOM)'', which is financially supported by the ``Nederlandse organisatie voor Wetenschappelijk Onderzoek (NWO)''.

\end{document}